\documentclass[aps]{revtex4}
\usepackage{graphicx}
\usepackage{float}

 \def\be {\begin{equation}}
\def\ee {\end{equation}}
\def\ba {\begin{eqnarray}}
\def\ea {\end{eqnarray}}

\def\e  {\epsilon}

\def\o  {\omega}

\def\p  {\pi}

\def\la {\label}
\def\le {\left}
\def\ri {\right}

\def\f {\frac}

\def\no {\noindent}
\def\bi {\begin{itemize}}
\def\ei {\end{itemize}}

\def\ra {\rangle}
\def\vs {\vspace}

\begin{document}
\title{Energy and Efficiency of Adiabatic Quantum Search Algorithms}
\author{Saurya Das
}
\email{saurya@math.unb.ca}
\affiliation{Dept. of Mathematics and Statistics, 
University of New Brunswick, 
Fredericton, N.B. - E3B 5A3, CANADA}
\author{Randy Kobes}
\email{randy@theory.uwinnipeg.ca}
\affiliation{ Physics Dept and Winnipeg Institute for Theoretical Physics,
The University of Winnipeg,
515 Portage Avenue,
Winnipeg, Manitoba R3B 2E9, CANADA}
\author{Gabor Kunstatter}
\email{g.kunstatter@uwinnipeg.ca}
\affiliation{ Physics Dept and Winnipeg Institute for Theoretical Physics,
The University of Winnipeg,
515 Portage Avenue,
Winnipeg, Manitoba R3B 2E9, Canada}

\begin{abstract}

We present the results of a detailed analysis of a general, 
unstructured adiabatic quantum search
of a data base of $N$ items. In particular we examine the effects on the computation time of adding
energy to the system. We find that by increasing the lowest eigenvalue of the time dependent
Hamiltonian {\it temporarily} to a maximum of $\propto \sqrt{N}$, it is possible to do the calculation in 
constant time. This leads us to derive the general theorem which provides the adiabatic analogue
of the $\sqrt{N}$ bound of conventional quantum searches. The result suggests that the action
associated with the oracle term in the time dependent Hamiltonian is a direct measure of the
resources required by the adiabatic quantum search.

\end{abstract}

\vs{.3cm}
\maketitle


\section{Introduction} 

The subject of quantum computation and quantum information theory has
attracted a great deal of attention in the recent past (for reviews, see
\cite{review1,review0,review}). This is mainly due to
the existence of many algorithms showing that the
principles of quantum  mechanics can be used to
greatly enhance the efficiency of solving specific computational
problems.  Notable among them are the factorization
algorithm due to Shor \cite{shor}, the algorithms of Deutsch and Jozsa
\cite{deutsch,dj} and  the data search algorithm due to Grover \cite{grover},
which is the subject of the present paper.

Assume that there are a total of $N$ items in a completely unstructured
database, out of which exactly one is marked (item $m$ say). The task
is to find that marked element in as few steps as possible.
Classically, on average $N/2$ steps are required, but
Grover was able to devise a quantum search algorithm which finds
the marked element in ${\cal O}(\sqrt{N})$ steps \cite{grover}. 
Moreover, this is the best that one can do since  Grover's
algorithm saturates the lower bound found by Bennett 
{\it et al} \cite{BBBV}
(henceforth referred to as the BBBV bound. See also \cite{fg,gg,zalka}).
Recently it was
suggested \cite{adia1,fg} that such gates can be replaced entirely
by a continuously time varying Hamiltonian which evolves a chosen
initial
quantum mechanical state directly to the required
marked item/state in an efficient way.
The details of this ``adiabatic'' approach are as follows: consider $N \equiv
2^n$ items in the
database, each associated with one vector in the complete orthonormal vector set
$\{ |i\ra, i=0,...,N-1\} $ in  the Hilbert space of
$n$ spin-$1/2$ objects. Assume that
the initial state of the quantum computer is the symmetric (normalized) state:
\be
| \psi_0 \ra = \f{1}{\sqrt{N}} \sum_{i=0}^{N-1} |i \ra~~.
\label{initial}
\ee
Let  $|m \ra$ denote the eigenstate associated with the marked item.
Now define the two Hermitian operators
\ba
H_0 &=& I - |\psi_0 \ra \langle \psi_0 |  \la{h0} \\
H_1 &=& I - |m \ra \langle m |  \la{h1}
\ea
whose ground states are $|\psi_0\ra$ and $|m\ra$, respectively,
and the time dependent Hamiltonian
\be
H (t) = (1-s(t)) H_0 + s(t) H_1~~.
\la{adham1}
\ee
In the above, $s(t)$ is some monotonically increasing function of
time $t$, subject to the conditions
\be
s(0)=0~~,~~s(T)=1~~.
\la{bc0}
\ee
where $T$ is the total computation time. Note that $H_1$, which projects
out the marked state, plays the role of the ``oracle'' of the Grover
algorithm. 

As long as the time evolution is slow enough, the adiabatic theorem guarantees
that the state $|\psi_0\rangle$ will evolve into the desired state $|m\rangle$
in
time $T$. In
Ref.\cite{rc} it was shown that the optimal running time for such an
adiabatic
quantum search goes as ${\cal{O}}(\sqrt{N})$, consistent with the BBBV bound.

The purpose of this letter is to present the results of a 
detailed and rigorous analysis of
the general form of this search algorithm.
The corresponding equations reduce exactly to a 
two-dimensional system which can easily be solved numerically for 
arbitrary $N$ without making any approximations. In principle, the 
generalized algorithm can be used to speed-up the search, but, 
as one might expect this comes at a cost: 
it is necessary to inject a large
amount of energy, albeit {\it temporarily} into the system. 
We illustrate this
mechanism with a specific example, and then derive the adiabatic
analogue of the BBBV bound and comment on its implications.

\section{A General Class of Models}

We begin by considering a time dependent Hamiltonian of the following 
general form:
\be
H(t)  =  f(t) H_0 + g(t) H_1~~,
\la{adham2}
\ee
where $H_0$ and $H_1$ are given by (\ref{h0}) and (\ref{h1}).  $f(t)$ and
$g(t)$ are arbitrary functions of time, subject
to the boundary conditions:
\ba
f(0) = 1~~&,&~~g(0)=0 \la{bc1} \\
\mbox{and}~~~f(T) = 0~~&,&~~g(T)=1 ~~. \la{bc2}
\ea
Once again, the system evolves from the initial state
$|\psi_0\ra$ to the marked state $|m\ra$ in time $T$. Note, however,
that unlike the usual algorithm, $f$ and $g$ are not necessarily monotonic (and
hence bounded above by one), nor do they not obey the constraint
$f+g=1$. This will play an important role later on.

It is a straightforward exercise to
find all $N$ eigenvalues $\{E_j(t)\}$  of the Hamiltonian $H(t)$ and their
corresponding eigenvectors $|E_j,t\rangle$. The two lowest eigenvalues are
\be
E_\pm (t) = \f{1}{2} \le[ (f+g) \pm
\sqrt{ (f-g)^2 + \f{4}{N} f g }   \ri]~~.
\la{diff1}
\ee
The higher energy eigenvalue is $(N-2)$ fold degenerate:
\be
E_i(t)= f+g ~~ ~~, ~~ ~~ i\neq\pm~~.
\ee
{}From the expression (\ref{diff1}) it follows that the eigenvalues $E_\pm$ are
non-degenerate as long as $f$ and $g$ are non-negative and
do not vanish simultaneously.

Without loss of generality we now write the solutions $|\psi(t)\rangle$ to the time dependent Schr\"odinger
equation in terms of the complete basis of energy eigenstates:
\be
|\psi(t)\rangle = \sum_{j=\pm,i} a_j(t) 
e^{-\frac{i}{\hbar}\int_0^t E_j(t') dt' } |E_j,t\rangle ~~.
\ee
The probability amplitudes
$a_\pm(t)$ for the two lowest energy eigenstates decouple from those of the higher excitations. They obey the simple equations:
\be
\dot{a}_\pm=F_\pm a_\mp ~~.
\label{exact}
\ee
In the above $ \dot{} \equiv d/dt$ and
\be
F_+ = - F^*_- =  \f{\sqrt{N-1}}{N}~\f{\dot{g}f-
{g}\dot{f}}{\omega^2}\exp\left(i\int^t_0\omega dt'\right) ~~,
\ee
and we have defined:
\be
\omega(t):=E_+(t)-E_-(t) =\sqrt{ (f-g)^2 + \f{4}{N} f g }~~.
\label{omega}
\ee
We have also used the
following  matrix elements:
\ba
 \le \langle E_+,t\le | \f{dH}{dt} \ri | E_-,t  \ri \ra
&=& \f{\sqrt{N-1}}{N}~\f{\dot{f}g -\dot{g}f}{\omega}
\la{mat1}\\
\le \langle E_\pm,t \le | \f{dH}{dt} \ri | E_i,t  \ri \ra &=&0 ~~, ~~
{}~~i\neq\pm ~~.
\label{off_diagonal}
\ea

According to the adiabatic theorem \cite{mess}, 
as used in \cite{adia1}, if there is a
small
number $\epsilon<<1$, such that for all $t$,
\be
 \le \langle E_+,t\le | \f{dH}{dt} \ri | E_-,t  \ri \ra \f{1}{\omega^2} \leq
\epsilon
\label{adia}
\ee
and if the system starts in the ground state $|\psi_0\ra $ of $H(t)$ at $t=0$,
then the probability of transition to the excited state, $|a_+(t)|^2$, will
not exceed a number
of order
$\epsilon^2$.
Note that for the theorem to hold,
the eigenvalues $E_\pm(t)$ must never cross. For a given desired computational
accuracy $\epsilon$, Eq.(\ref{adia})
limits the rate at which the Hamiltonian can evolve, and therefore puts a lower
bound on the total running time $T$. In order to calculate the bound,
one needs more information about the form of functions $f(t)$ and $g(t)$.
For example, if $f(t)=1-s(t)$ and $g(t)= s(t)$, as in
Eq.(\ref{adham1}),
the total running  time $T$ can be minimized by
 choosing $ds/dt$ to saturate the bound (\ref{adia}) at each instant in time.
This results in a running time of order $\sqrt{N}$ \cite{rc}, consistent with
the BBBV bound \cite{BBBV}.

\section{A Specific Example}

We now illustrate with a specific example that more 
general choices for the functions
$f(t)$ and
$g(t)$ allow the computation
to be performed considerably more rapidly.
Consider $f(t)$ and $g(t)$
to be  quadratic functions of $s(t)$:
\be
f (s) = 1 - s  + As(1-s)~~,~~
g(s) = s + As(1-s) ~~~(0\leq s \leq 1)~~,  \la{quad1}
\ee
where $A$ is a constant and $s(t)$ is again a function of time that varies
monotonically
between 0 and 1, so that the required boundary conditions (\ref{bc1}-\ref{bc2})
are satisfied. Note that $f$ and $g$ are not monotonic in this case. Their peak
values are
$A/4+1/2 \mp 1/4A$ respectively (occurring at $s=1/2(1\mp 1/A)$ respectively).
The ``adiabaticity condition'' (\ref{adia}) in this case becomes
(we use $d/dt = (ds/dt)~d/ds$)
\ba
\f{ds}{dt}\f{\sqrt{N-1}}{N} \f{ 1+A(1 - 2s+2s^2)  }{ \le[ (1-2s) +
\f{4}{N}
s(1-s) \le[ 1 + A - s(1-s) A^2 \ri]  \ri]^{3/2}}\leq\epsilon~~.
\label{gen_adiab}
\ea

For a system satisfying the constraint (\ref{gen_adiab}), the minimum
computation time can again be found
by adjusting $s(t)$ so that the inequality is saturated at all times.
This yields a minimum running time of:
\be
 T_{min} = \f{1}{\e} \f{\sqrt{N-1}}{N}
\int_0^1 ds\f{\le|1+A(1 - 2s+2s^2) \ri| }{ \le[ (1-2s) +
\f{4}{N}
s(1-s) \le[ 1 + A - s(1-s) A^2 \ri]  \ri]^{3/2}}~~.
\label{minimum1}
\ee
If $A=0$, the result of
\cite{adia1} is reproduced. However, by choosing 
$A$ appropriately, considerably lower running
times can be achieved. For example, if $A = \sqrt{N} \gg1$ we get :
\ba
T_{min}&=& \f{1}{\e} \le( 1 + \f{\p}{2} \ri) +{\cal
O}(\f{1}{\sqrt{N}})\la{time4} ~~.
\label{minimum}
\ea
The limit (\ref{minimum}) can be obtained by Taylor expanding the 
integrand in (\ref{minimum1}) about $N=\infty$. This is possible
for the choice $A=\sqrt{N}$ because the denominator does not vanish
anywhere when $N=\infty$. One has to be considerably more careful in deriving the limit for other values of $A$. We have also verified (\ref{minimum}) numerically.

Thus we see that the data search can be completed in a constant time,
independent of $N$. The cost of this improvement is that the energy of the
system gets very large for a finite time before returning to zero at the end of
the computation. In particular, for the choices of $f$ and $g$ in
(\ref{quad1}), $E_-(s)$ reaches a maximum of $(A+1)/4$ (for $N\gg 1$)  at
$s=1/2$, so that if $A=\sqrt{N}$, the maximum energy grows with $\sqrt{N}$ as
well.

\begin{figure}[H]
\begin{center}
\includegraphics[width=3.375in]{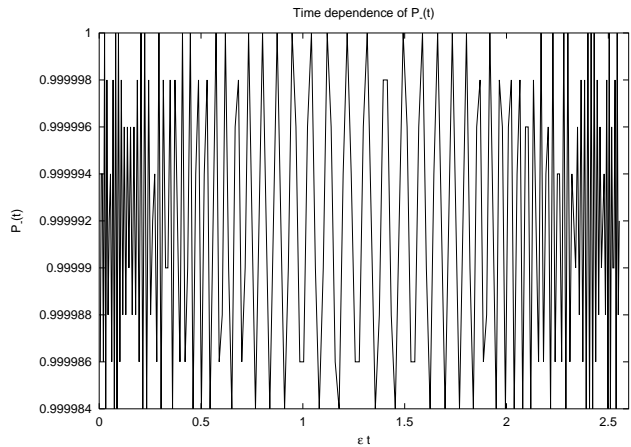}
\end{center}
\caption{Numerical solution for $P_-(t)\equiv |a_-(t)|^2$ for
$N=100,000$ and $\epsilon=0.002$ with $A=\sqrt{N}$.}
\label{apm}
\end{figure}

By numerically solving (\ref{exact}), it can be confirmed that
the quantum computer indeed remains in the ground state at all times. 
One can explicitly calculate the
transition probability $P_-(t)=|a_-(t)|^2$ as a function of time (by the
equations of motion (\ref{exact}), the
corresponding $P_+(t)=|a_+(t)|^2$ is guaranteed to be $1-P_-(t)$).  
Figure \ref{apm}
illustrates a sample solution with $A=\sqrt{N}$ and $s(t)$ chosen to saturate
the inequality
(\ref{gen_adiab}) at all times $t$. The
transition
probability $|a_-|^2$ remains above $\sim(1-\epsilon^2)$ and the graph 
is plotted till the end of running time 
$\e t = 1+\pi/2 \approx 2.57$. 

\par

\par\noindent
We have explored numerically other choices of $A$.
For $A\sim N^\alpha$, where $\alpha \ge 0.5$, 
the calculational running time 
approaches a constant for large $N$, with the constant of 
proportionality decreasing as $\alpha$ increases.
A systematic analysis of this problem is in progress \cite{dkk3}, 
but in light of this and the
discussion below, it is interesting to speculate
what happens to the running time for
a non-polynomial choice such as $A\sim e^{N}$.


\section{Adiabatic Version of the BBBV Bound}

At this stage it is natural to ask what are the most general conditions 
that allow a speed-up of the search algorithm. This question is answered in
the following theorem, which
generalizes the results of
\cite{BBBV,fg,gg,zalka,rc}.
\begin{quote}
\underline{THEOREM:} Suppose there exists an unstructured adiabatic
quantum search algorithm that can successfully
evolve the initial state (\ref{initial}) to any one of a complete set
of orthogonal states $\{|m \ra\}$  via a
Hamiltonian containing an oracle term of the form $g(t)|m><m|$. Then the time 
$T$ for the
computation is
bounded below by the relationship:
\be
{1\over \hbar}\int^T_0 g(t)dt\geq \frac{ k\sqrt{N}}{4}
\label{bound}
\ee
in the limit that $N$ is large ($k$ is a constant of order unity).
\end{quote}
This inequality is virtually identical to the one obtained by \cite{rc}. 
The one important difference is that we do not restrict $g(t)$ to be
less than one, nor do we impose the constraint $f(t)+g(t)=1$.
Details of the proof can be found in \cite{rc} so we will 
not repeat the details, merely outline the important ingredients. 
The key is to decompose the Hamiltonian into two parts:
\be
H (t) = H_{2m} (t) + H_1(t)~~,
\ee
where %
\ba
H_{2m} (t) &=& - g(t) |m\ra \langle m |  ~~.
\ea
 contains all the dependence on the needle state $|m\ra$
and
$H_1(t)$ is an arbitrary term independent of the state
$|m \ra$ (in particular, for the Hamiltonian (\ref{adham2}), 
$H_1(t) = f(t) + g(t) - f(t) |\psi_0\ra\langle \psi_0|$).
Note, however, that $H_1(t)$ is completely general, apart from
the restriction that it not depend explicitly on the
state $|m\ra$.

Consider two computers at time $t$, evolving to states
$|m\ra$ and $|m'\ra$
respectively, represented by the wavefunctions
$|\psi_m,t\ra$ and $|\psi_{m'},t\ra$
respectively
subject to the boundary conditions:
\ba
|\psi_m, 0 \ra &=& |\psi_{m'}, 0 \ra = |\psi_0 \ra  
\la{bc11} \\
|\psi_m, T \ra &=& |m\ra ~~~;~~~ 
|\psi_{m'}, T \ra = |m' \ra~~~. \la{bc22}
\ea
In terms of the specific model we considered earlier, this
requires $f(T)=0$ and $g(T)=1$.
Following the steps in \cite{rc} it is straightforward to
obtain the inequality:
\be
\sum_{m,m'} \le[ 1 - |\langle \psi_m,T|\psi_{m'},T \ra
|^2 \ri] 
\leq {4 N^{3/2}\over \hbar}  \int_0^T g(t) dt~~.
\la{time5}
\ee
The assumed orthogonality of the final states at time $T$ implies
that
\be
1 - |\langle \psi_m,T|\psi_{m'},T\ra|^2 \geq
k~~~,\forall m \neq m'
\ee
where $k$ is a number of order 1.
Summing over $m$ and $m'$ in 
Eq.(\ref{time5})  for $N \gg 1$ yields Eq.(\ref{bound}) as desired. Again
we stress the Eq.(\ref{bound}) is virtually identical to the result 
obtained in \cite{rc}. The key difference concerns our relaxed assumptions
about the functions $f$ and $g$ in the Hamiltonian.

Eq.(\ref{bound}) shows that the BBBV bound can only be beaten if the mean
value of $g(t)$
over the time $T$ grows as some power of
$N$. If this power is $1/2$ then the above theorem states that the
running time $T$ may be bounded by a constant independent
of $N$, as in the example quoted in this paper. Note that this
result is a generalization of a similar bound obtained in \cite{rc} for
the special case given by (\ref{adham1}), in which $g(t)$ 
never exceeds order unity. It also is a natural extension of the result
presented by Farhi and Gutmann in \cite{fg} for constant $g$.

\section{Conclusions} 

In summary, we have presented the results of an analysis of a 
generalized adiabatic quantum search algorithm. 
The corresponding  Schr\"odinger equation was shown to 
reduce exactly to a two dimensional system for arbitrary $N$. 
We derived the adiabatic analogue of the BBBV bound. 
Our theorem shows that the optimal speed normally associated with
Grover's search algorithm can be
improved in this framework by a suitable choice of the
time-dependent Hamiltonian. As one might expect from dimensional 
grounds this speed-up requires an increase in energy, at least 
temporarily. However, it should be emphasized that
it is not the total Hamiltonian that needs to be increased, 
only the coefficient in front of the
oracle term. In principle this leaves open the possibility of speeding up
the search while keeping the ground state energy of the system small. 
Another way to keep the ground state energy zero would be to 
use a new Hamiltonian obtained from (\ref{adham2}) by subtracting
the term $E_-(t) I$ from the latter. Note that although the
resultant Hamiltonian cannot be written in the form $fH_0 + gH_1$,
since $E_-(0)=0=E_-(T)$, it would still evolve the initial state
to $|m\ra$. Moreover, the `gap' $\o(t)$, as well as the matrix element
(\ref{mat1}) will remain intact, implying that the running time will still be 
a constant, given by (\ref{minimum}).
In any case, our analysis suggests that the physical
quantity ${1\over \hbar}\int _0^T g(t) dt $ provides an adiabatic analogue
of resources required for the unstructured search, just as the number of
operations does for the conventional quantum search. 

It must be emphasized that the system considered here is highly idealized. 
The Hamiltonian is non-local in the sense that $H_0$ and $H_1$ 
require all qubits to be coupled simultaneously. 
It is therefore not clear how to implement such a
Hamiltonian in a realistic physical system. One should therefore investigate
the circumstances
under which one can find simpler, more local, 
Hamiltonians that have the same ground states and hence can 
be used as the basis of a realistic adiabatic quantum search algorithm. 
This issue will be the subject of a separate publication\cite{adkkz2}.


\vs{.4cm}
\no
{\bf Acknowledgements}

\no
We would like to thank D. Ahrensmeier, J. Currie,
 R. Laflamme, V. Linek and
H. Zaraket for useful
discussions and encouragement
at various stages during this work. G.K. also thanks
N. Cerf and J. Roland for helpful input.
We are grateful to A. Childs, E. Farhi, S. Gutmann, H. Ollivier and
D. Poulin for comments on an earlier version of this paper.
This work was supported in part by the Natural Sciences and
Engineering Research Council of Canada.


\end{document}